\documentclass[iop]{emulateapj}
\usepackage{lscape}
\shorttitle{M-Z relation for LGRB Host Galaxies}
\shortauthors{Levesque et al.}
\slugcomment{DRAFT \today}

\begin{document}

\title{The Host Galaxies of Gamma-Ray Bursts II: A Mass-Metallicity Relation for Long-Duration Gamma-Ray Burst Host Galaxies}
\author{Emily M. Levesque$^{1,2,3}$}
\email{Emily.Levesque@colorado.edu}
\author{Lisa J. Kewley\footnotemark[1]}
\author{Edo Berger\footnotemark[2]}
\author{H. Jabran Zahid\footnotemark[1]}

\begin{abstract}
We present a statistically robust mass-metallicity relation for long-duration gamma-ray burst (LGRB) host galaxies at $z < 1$. By comparing the LGRB host mass-metallicity relation to samples representative of the general star-forming galaxy population, we conclude that LGRBs occur in host galaxies with lower metallicities than the general population, and that this trend extends to $z \sim 1$, with an average offset of $-0.42 \pm 0.18$ from the {\it M-Z} relation for star-forming galaxies. Our sample in this work includes new spectroscopic data for 6 LGRB host galaxies obtained at the Keck and Magellan telecopes, as well as 2 new host galaxies from the literature. Combined with data from our previous work, this yields a total sample 6 LGRB host galaxies at $z < 0.3$ and 10 host galaxies at $0.3 < z < 1$. We have determined a number of interstellar medium properties for our host galaxies using optical emission-line diagnostics, including metallicity, ionization parameter, young stellar population age, and star formation rate. Across our full sample of 16 LGRB hosts we find an average metallicity of log(O/H) + 12 = 8.4 $\pm$ 0.3. Notably, we also measure a comparatively high metallicity of log(O/H) + 12 = 8.83 $\pm$ 0.1 for the $z = 0.296$ host galaxy of GRB 050826. We also determine stellar masses ($M_{\star}$) for our LGRB host galaxy sample, finding a mean stellar mass of log($M_{\star}/M_{\odot}$) = 9.25$^{+0.19}_{-0.23}$.
\end{abstract}

\footnotetext[1]{Institute for Astronomy, University of Hawaii, 2680 Woodlawn Dr., Honolulu, HI 96822}
\footnotetext[2]{Smithsonian Astrophysical Observatory, 60 Garden St., MS-20, Cambridge, MA 02138}
\footnotetext[3]{Current address: CASA, Department of Astrophysical and Planetary Sciences, University of Colorado, 389-UCB, Boulder, CO 80309}

\section{Introduction}
\label{Sec-intro}
Long-duration gamma-ray bursts (LGRBs), thought to be generated during the core-collapse of massive stars (Woosley 1993), are among the most energetic phenomena observed in the universe. LGRBs are typically associated with star-forming host environments and young massive star progenitors. As a result, a number of studies cite these explosive events as potential unbiased tracers of the star formation and metallicity history of the universe at high redshifts (e.g., Bloom et al.\ 2002, Fynbo et al.\ 2006a, Chary et al.\ 2007, Savaglio et al.\ 2009). However, several recent studies have uncovered a connection between LGRBs and low-metallicity galaxies. Such a trend could potentially challenge the use of these phenomena as tracers of star formation in normal galaxies at large look-back times.

Much of the previous work on LGRBs and their metallicities has focused on a comparison with the standard luminosity-metallicity ({\it L-Z}) relation for star-forming galaxies. Stanek et al.\ (2006) examined the host galaxies of 5 $z < 0.3$ LGRBs and found that their metallicities were lower than equally-luminous dwarf irregular galaxies. Kewley et al.\ (2007) noted that these $z < 0.3$ LGRBs occupied the same position on the {\it L-Z} diagram as low-metallicity galaxies, falling below the general relation for dwarf irregular galaxies. Modjaz et al.\ (2008) similarly found that LGRB host galaxies had lower metallicities than the host galaxies of nearby broad-lined Type Ic supernovae. Fynbo et al.\ (2008) adopt a theoretical {\it L-Z} relation for use with higher-redshift ($z \sim 3$) observations of LGRB hosts, suggesting that these higher-redshift hosts may be good representative samples of higher-redshift star-forming galaxies.

The mass-metallicty ({\it M-Z}) relation is cited as the fundamental property that drives the observed {\it L-Z} relation. While luminosity is often adopted as a proxy for stellar mass, a galaxy's luminosity is also extremely dependent on star formation rate (SFR) and star formation history as well as metallicity, and thus does not effectively isolate stellar mass as a parameter. The {\it M-Z} relation for nearby galaxies may be attributable to the larger neutral gas fractions and more efficient stripping of heavy elements by galactic winds in lower-mass galaxies (McGaugh \& de Blok 1997, Bell \& de Jong 2000, Boselli et al.\ 2001, Garnett 2002, Tremonti et al.\ 2004), though this may not be the dominant effect driving the {\it M-Z} relation at higher redshifts (see Zahid et al.\ 2010). 

Work on the {\it M-Z} relation dates back to Lequeux et al.\ (1979), who found a positive correlation between mass and metallicity that agreed with model predictions for six nearby irregular galaxies. More recently, Tremonti et al.\ (2004) found a {\it M-Z} relation for $\sim$53,000 nearby ($z < 0.3$) star-forming galaxies from the Sloan Digital Sky Survey. Savaglio et al.\ (2005) found that this correlation extended to higher redshifts, based on observations of galaxies from the Gemini Deep Deep Survey (GDDS; Abraham et al.\ 2004) at $0.4 < z < 1$. Erb et al.\ (2006) measured a monotonic {\it M-Z} relation for galaxies at a mean redshift of $z \sim 2$, and found that this relation was offset from the local {\it M-Z} relation by $\sim$0.3 dex, with galaxies of a given stellar mass having lower metallicities at higher redshifts.

Determining an {\it M-Z} relation for LGRB host galaxies, and comparing this relation to the general galaxy population, is critical. A clearer understanding of the potential metallicity trend in LGRB hosts is key to determining whether they are potential tracers of star formation in the higher-redshift universe, and the {\it M-Z} relation is the best means of isolating the effects of metallicity. If LGRB hosts are found to consistently lie below the general {\it M-Z} relation, this result would suggest that LGRBs do occur preferentially in low-metallicity galaxies.

However, it is important to note that a low-metallicity trend would not entirely preclude the use of LGRBs as tracers of star formation in the high-redshift universe. The metallicity of the star-forming galaxy population is known to evolve with redshift; galaxies at $z \ge$ 1 are less enriched and have lower metallicities on average (e.g., Kobulnicky \& Kewley 2004, Shapley et al.\ 2004, Erb et al.\ 2006, Chary et al.\ 2007, Dav\`{e} \& Oppenheimer 2007, Liu et al.\ 2008). Furthermore, the {\it M-Z} relation is known to evolve with redshift (Savaglio et al.\ 2005). Based on the suggested cut-off metallicity for $z < 0.25$ LGRB host galaxies (log(O/H) + 12 = 8.66 from Modjaz et al.\ 2008), Kocevski et al.\ (2009) derive a redshift-dependent upper limit for the stellar mass of a galaxy that can efficiently produce a LGRB. They predict that the peak of the GRB host stellar mass distribution and the stellar mass where SFR peaks should coincide at $z \sim 2$, suggesting that at higher redshifts LGRBs may indeed be trace the general star-forming galaxy population (this equality point moves to lower redshifts with a higher metallicity cutoff).

In Paper I, we presented the first results of our ongoing uniform rest-frame optical spectroscopic survey of nearby ($z < 1$) LGRB host galaxies. We determined ISM properties, including metallicity, star formation rate (SFR), and the age of the young stellar population, for 10 LGRB host galaxies. This work offered additional strong evidence for a low-metallicity trend in LGRB host galaxies. By comparing the Paper I host galaxies to a variety of star-forming galaxy samples from the general population on the ({\it L-Z}) diagram, we found that at $z < 0.3$ the metallicity offset between LGRB host galaxies and the general population was significant. However, at higher redshifts ($0.3 < z < 1$) this offset was less robust. This comparison would benefit greatly from comparing LGRB host galaxies to the general population using the more fundamental {\it M-Z} relation rather than the {\it L-Z} relation.

In this paper, we continue our survey of LGRB host galaxies from Paper I, detailing the observations and host data taken from the literature (Section 2) as well as our means of determining the host interstellar medium (ISM) properties (Section 3). We have determined stellar masses for our host galaxies, and use this data to construct a {\it M-Z} relation for our LGRB host sample, which we compare to general star-forming galaxy populations (Section 4). Finally, we discuss the implications of our results and the impact on future work in this area (Section 5). Throughout this work we assume a cosmology of $H_0 = 70$ km s$^{-1}$ Mpc$^{-1}$, $\Omega_m = 0.3$, and $\Omega_\Lambda = 0.7$.

\section{Observations}
\label{data}
\subsection{Nearby LGRB Host Galaxy Survey}
We are conducting a uniform rest-frame optical spectroscopic survey of nearby LGRB host galaxies using the Keck telescopes on Mauna Kea and the Magellan telescopes at Las Campanas Observatory. The sample included in this survey was compiled from the GHostS database (Savaglio et al.\ 2006) and the Gamma-Ray Burst Coordinates Network. We restrict our sample to confirmed host galaxies of long-duration ($>$ 2 s) GRBs with redshifts of $z < 1$, allowing us to obtain rest-frame optical spectra from 3000-7000\AA\ using optical and near-infrared observations. Eight galaxies from this ongoing survey were presented in Paper I; observations of four new galaxies, along with additional observations of two hosts from Paper I, and data from two new host galaxies in the literature, are included here.

\subsubsection{Keck: GRBs 980703, 991208, 010921, 020819, and 070612A}
We obtained spectra of 5 LGRB host galaxies using the Near Infrared Spectrograph (NIRSPEC) and the Low-Resolution Imaging Spectrograph (LRIS) on the Keck telescopes at Mauna Kea. We observed the host galaxies of GRB 980703, GRB 010921, GRB 020819, and GRB 070612A in the rest-frame optical using LRIS on 18-19 November 2009. We also obtained observations of the GRB 070612A host on 3 November 2009 and the GRB 991208 host on 2 May 2010 using NIRSPEC to detect the H$\alpha$ and [NII]$\lambda$6584 emission features shifted into the near-infrared for these $z > 0.6$ hosts.

For the LRIS observations, we used the long 1" slitmask. For calibration, we obtained internal flat fields and comparison lamp spectra with the standard Hg, Ne, Ar, Cd, and Zn lamp setup available at LRIS. We flux-calibrate the host spectra using contemporaneous observations of spectrophotometric standards. Our observation details are given in Table 1.

These host galaxies are quite dim ($V \sim$ 20 to 25 mag). To ensure that we successfully placed such dim targets on the slit, we first centered on a nearby bright star. We would then rotate the slit to a position angle that would place both the bright star and the host galaxy on the slit, and nod along the slit to ensure that we observed spectra of both objects. This approach also allowed the bright spectrum to be used as a trace when extracting the dim host spectrum during data reduction. As a result of this method, we did not observe the host galaxies at the parallactic angle.

The host galaxy of GRB 010921 had been previously observed with LRIS and published in Paper I. We obtained additional LRIS observations in November of 2009 using the newly updated red side of the LRIS detector, which offered improved sensitivity in the $>$9500\AA\ regime and made it possible for us to observe the key diagnostic emission features H$\alpha$ and [NII]$\lambda$6584 (for which we determine an upper limit) at the host redshift of $z = 0.451$. This allowed us to further refine the metallicity and other ISM properties of this host. The LRIS observations of GRB 020819 described here were published in Levesque et al.\ 2010b; we include them in this discussion for completeness.

For the NIRSPEC observations of GRB 991208 and 070612A, we used the 42"$\times$0.76" slit with the low-resolution grating. We observed internal flatfields and darks for calibration purposes. The host of GRB 991208 was observed using the NIRSPEC-2 filter in six 900 s exposures. GRB 070612A was observed using the NIRSPEC-1 filter, in a single 900 s exposure with one coadd. The host of GRB 070612A is sufficiently bright ($R \sim 21.4$, D'Avanzo et al.\ 2007) that we could center on it directly, rather than employing the technique described above for observing faint objects with LRIS using nearby bright stars placed on the slit. For both of these hosts we detect the H$\alpha$ emission feature and place an upper limit on the relative flux of the [NII]$\lambda$6584 emission line. In the case of GRB 991208, this allowed us to further refine the ISM properties published in Paper I based on previous LRIS observations.

\subsubsection{Magellan: GRB 050826}
The host galaxy of GRB 050826 was observed twice using LDSS3 at the Clay 6.5m Magellan telescope at Las Campanas Observatory. Two 1800 second exposures of the host galaxy were taken on 6 January 2006, using the VPH-Red grism and an OG590 blocking filter and including strong detections of the H$\alpha$, [NII]$\lambda$6584, and [SII]$\lambda\lambda$6717,6731 emission features. An additional two 1800 second exposures were taken of the host on 14 January 2008, using the VPH-All grism and a 1" slit to include full spectral coverage from the H$\alpha$ feature down to the [OII]$\lambda$3727 features. The observations were taken at the parallactic angle. Internal flatfields, along with lamp spectra of He, Ne, and Ar, were observed calibration purposes. Contemporaneous observations of the spectrophotometric standard LTT 3864 (Hamuy et al.\ 1994) were used for flux calibration.

\subsubsection{Published LGRB Host Spectra: GRB 030528 and GRB 050824}
Emission line fluxes for GRB 030528 and GRB 050824 were taken from previous observations published in Rau et al.\ (2005), and Sollerman et al.\ (2007) respectively. Rau et al.\ (2005) observed the host of GRB 030528 using the Focal Reducer and low-dispersion Spectrograph 2 (FORS2) at the 8.2m Very Large Telescope (VLT) on 12 April 2005 and 6 May 2005. Sollerman et al.\ (2007) observed the afterglow and host of GRB 050824 using FORS2 at the VLT on 26-27 August 2005; the afterglow contribution to the emission line fluxes observed in this host is assumed to be negligible.

\subsubsection{Data Reduction}
We reduced and analyzed the LGRB host galaxy data from LRIS and NIRSPEC using IRAF\footnotemark[1]. \footnotetext[1]{IRAF is distributed by NOAO, which is operated by AURA, Inc., under cooperative agreement with the NSF.} For the LRIS observations, we used the \texttt{lrisbias} IRAF task distributed by the W. M. Keck Observatories to subtract the overscan from the LRIS images. We applied a flatfield correction based on our internal lamp flats. The spectra were extracted using an optimal extraction algorithm, with deviant pixels identified and rejected based upon the assumption of a smoothly varying profile. Wavelength calibration and flux calibration were performed using our arc lamp spectra and observations of spectrophotometric standards. We measured the emission line fluxes in these spectra using the IRAF task \texttt{splot} in the \texttt{kpnoslit} package to fit Gaussians to the line profiles.

To reduce the NIRSPEC data, we used the \texttt{wmkonspec} data reduction package distributed by the W. M. Keck Observatories. We used the \texttt{xdistcor}, \texttt{ydistcor}, and \texttt{mktracer} IRAF tasks to correct for x- and y-axis distortion in the observed spectrum. The spectra were extracted using the same algorithm applied to the LRIS data; in addition, a sky spectrum was extracted, and we used the \texttt{skyplot} task in the \texttt{wmkonspec} package to generate a comparison spectrum that could be used in conjunction with the extracted sky spectrum for wavelength calibration. 

The raw fluxes that we measure for each of our observed host galaxies are given in Table 2.

\section{Determining Host Galaxy Properties}
\label{ISM}
To derive ISM properties for these LGRB host galaxies, we adopt the same diagnostics and procedures outlined in Paper I. We begin by correcting the observed line fluxes for the hosts using E($B-V$), which we determine based on the fluxes of the Balmer lines (H$\alpha$ and H$\beta$ where available, H$\beta$ and H$\gamma$ otherwise) and the Cardelli et al.\ (1989) reddening law. Using these dereddened line fluxes, we determine metallicities, ionization parameters, young stellar population ages, and star formation rates (SFR) for the host galaxies.

\subsection{Metallicity}
We determine metallicities using the theoretical ([OIII]$\lambda$5007 + [OIII]$\lambda$4959 + [OII]$\lambda$3727)/H$\beta$ (R$_{23}$) diagnostic originally presented in Kewley \& Dopita (2002) and refined in Kobulnicky \& Kewley (2004). Since this diagnostic is double-valued, we apply several tests to determine whether each galaxy's metallicity should be determined by the ``lower"- or ``upper"-brach diagnostic equations. The [NII]$\lambda$6584/[OII]$\lambda$3727 ratio was used to differentiate between the upper and lower branches of the R$_{23}$ diagnostic where available (for the hosts of GRB 020819 and GRB 050826), with log([NII]/[OII]) $< -1.2$ indicating lower branch and log([NII]/[OII]) $> -1.2$ indicating upper branch (Kewley \& Ellison 2008). The [NII]$\lambda$6584/H$\alpha$ ratio provided an alternate means of determining the diagnostic branch where [NII]/[OII] was not available (for the hosts of GRB 010921, GRB 050826, and GRB 070612A). With this criterion, log([NII]/H$\alpha$) $< -1.3$ indicates lower branch and log([NII]/H$\alpha$) $> -1.1$ indicates upper branch, leaving an indeterminate range of values in between (Kewley \& Ellison 2008).

The R$_{23}$ values for the hosts of GRB 030528 and GRB 050824 fall on the ``turn-over" of the Kobulnicky \& Kewley (2004) diagnostic, corresponding to a metallicity of log(O/H) + 12 $\sim$ 8.4. For the host galaxy of GRB 980703 we could not determine whether the galaxy has a lower- or upper-branch metallicity in the Kobulnicky \& Kewley (2004) diagnostic, since we have no detection or upper limit for the H$\alpha$ and [NII] $\lambda$6584 features in this $z = 0.966$ host. We find a similar double-valued metallicity for one host galaxy from Paper I (GRB 020405). We therefore give metallicites determined from both the lower- and upper-branch equations for these galaxies in Table 3.

For comparison where possible, we also determined metallicities using the Pettini \& Pagel (2004) relation between log(([OIII]$\lambda$5007/H$\beta$)/([NII]$\lambda$6584/H$\alpha$)) ($O3N2$) and metallicity, where log(O/H) + 12 $ = 8.73 - 0.32 \times O3N2$. This method, based on a calibration of nebular HII region metallicities, is known to yield systematically lower metallicities than theoretical methods based on photoionization models (see Kewley \& Ellison 2008).

Combining the LGRB host galaxies examined in this work with the sample from Paper I we find an average R$_{23}$ metallicity for these 16 $z < 1$ LGRB host galaxies of log(O/H) + 12 = 8.4 $\pm$ 0.3. For the eight host galaxies in our sample with Pettini \& Pagel (2004) metallicities, we find an average $O3N2$ metallicity of log(O/H) + 12 = 8.3 $\pm$ 0.3.

\subsection{Ionization Parameter}
Here we define the ionization parameter $q$ in cm s$^{-1}$ as the maximum velocity possible for an ionization front being driven by the local radiation field, where $q$ relates to the dimensionless ionization parameter ($\mathcal{U}$) by $\mathcal{U} \equiv q/c$. The value of $q$ itself is calculated for the inner surface of the nebula. We determine $q$ using the Kewley \& Dopita (2002) [OIII]/[OII]-$q$ relation. For the combined sample of LGRB host galaxies in this work and Paper I, we find an average log($q$) = 7.7 $\pm$ 0.3 dex.

\subsection{Young Stellar Population Ages}
The young stellar population ages for our LGRB host galaxies were determined using the metallicity-dependent polynomial relation between H$\beta$ equivalent width and age, published in Paper I and based on the models of Schaerer \& Vacca (1998). This relation assumes a zero-age instantaneous burst star formation history for the host galaxy (Copetti et al.\ 1986). For our sample of LGRB host galaxies included here and in Paper I, we find an average age of 5.6 $\pm$ 1.2 Myr, an age which corresponds to the Wolf-Rayet evolutionary phase in young stellar populations (e.g. Schaller et al.\ 1992; Schaerer et al.\ 1993a, 1993b; Charbonnel et al.\ 1993).

\subsection{Star Formation Rate}
The H$\alpha$ emission feature is currently considered the most reliable optical tracer of star formation rate (SFR) in a galaxy, due to the direct scaling of H$\alpha$ with the total ionizing flux of newly formed stars in the ionizing nebulae (e.g. Kennicutt 1998, Kewley et al.\ 2004). The [OII]$\lambda$3727 feature is also a useful means of determining SFR for galaxies. However, this latter feature is strongly dependent on the  chemical abundance of the galaxy, requiring that relations between [OII] and SFR take metallicity into consideration (Kewley et al.\ 2004).

For the host galaxies of GRB 020819 and GRB 050826, we were able to determine SFR based on the H$\alpha$ relation of Kennicutt (1998), SFR(M$_{\odot}$/yr) = (7.9 $\times$ 10$^{-42}$) $\times$ L(H$\alpha$), where possible. For the remainder of our hosts, where H$\alpha$ fluxes were not available, we instead adopted the metallicity-dependent SFR relation for [OII]$\lambda$3727 luminosities from Kewley et al.\ (2004). In Paper I we found that the SFRs for LGRB host galaxies spanned an exceptionally wide range, from 0.03M$_{\odot}$/yr for the host of GRB 060218 to the remarkably high 271M$_{\odot}$/yr for the host of GRB 051022 (the latter host may possibly be a merging system; see Graham et al.\ 2009). The SFRs of the LGRB hosts in this paper all fall within this wide range, consistent with galaxies undergoing active star formation.

\subsection{Stellar Mass}
We have estimated stellar masses for all of the LGRB host galaxies in this paper, as well as those presented in Paper I. Using the {\it Le Phare}\footnotetext{http://www.cfht.hawaii.edu/$\sim$arnouts/LEPHARE/cfht\_lephare/\\lephare.html} code developed by S. Arnouts \& O. Ilbert, we fit multiband photometry for the host galaxies taken from Savaglio et al.\ (2009) with stellar population synthesis models generated from the Bruzual \& Charlot (2003) synthetic stellar templates and the initial mass function (IMF) of Chabrier (2003). For these mass determinations we adopt the extinction law of Calzetti et al.\ (2000). This fitting yields a stellar mass probability distribution for each galaxy, and we take the median of the distribution as an estimate of the final stellar mass. For a more detailed discussion of how stellar mass is estimated using the {\it Le Phare} code, see Ilbert et al.\ (2009). This differs from the method of determining stellar masses described in Savaglio et al.\ (2009), who adopt different stellar population synthesis models and simulate contributions from both old and young stellar populations (for more discussion see Glazebrook et al.\ 2004). We find that our stellar mass determinations generally agree with the values from Savaglio et al.\ (2009) to within the errors. For the complete sample of LGRB host galaxies, we find a mean stellar mass of log($M_{\star}/M_{\odot}$) = 9.25$^{+0.19}_{-0.23}$.

The ISM properties and stellar masses determined for our LGRB host galaxies are given in Table 3. For a detailed discussion of the parameters we determined for each host galaxy, see the Appendix.
 
\section{The Mass-Metallicity Relation for LGRB Hosts}
In recent years, stellar masses for LGRB host galaxies have been examined in some detail. Castro Cer\'{o}n et al.\ (2006) estimated the stellar masses for 6 LGRB host galaxies at $z \sim 1$ using $K$ band fluxes; in Castro Cer\'{o}n et al.\ (2008) this work was extended to $K$-band stellar mass estimates for 16 LGRB hosts and upper limits on stellar mass for an additional 14 hosts. The LGRB hosts were all found to be low-mass star-forming systems with 7 $<$ log($M_{\star}/M_{\odot}$) $<$ 11 (median log($M_{\star}/M_{\odot}$) = 9.7), at $0.009 < z < 2.66$. Castro Cer\'{o}n et al.\ (2008) found that the median stellar mass for LGRB hosts was lower than the median of galaxies from GDDS, and did not detect any intrinsic evolution of stellar mass with redshift.

Savaglio et al.\ (2009) determined stellar masses and metallicities for a number of LGRB hosts, as well as short-duration GRB hosts, finding an average log($M_{\star}/M_{\odot}$) = 9.3 for the full sample. Based on these data they do not find any {\it M-Z} relation for 16 GRB hosts with measured metallicities. Savaglio et al.\ (2009) also see no metallicity offset when comparing the GRB host sample to local star-forming dwarf galaxies from Lee et al.\ (2006) and higher-redshift galaxies from GDDS. However, these comparisons are conducted using metallicities from several different diagnostics, which are known to show considerable disagreements and offsets in their results and require careful polynomial conversions if values from different calibrations are to be compared (see discussion Kewley \& Ellison 2008). Savaglio et al.\ (2009) also note that the metallicities are poorly constrained for 9 of the 16 GRB hosts in their sample.  Finally, the Savaglio et al.\ (2009) comparison includes several host galaxies of GRBs that they classify as short-duration (GRB 051221, GRB 050416; although see Soderberg et al.\ 2007) as well as the unusual GRB 060505, a burst whose phenomenological classification remains unclear (e.g. Fynbo et al.\ 2006b, Levesque \& Kewley 2007, Ofek et al.\ 2007, McBreen et al.\ 2008, Th\"{o}ne et al.\ 2008). Short-duration GRBs are thought to be phenomenologically distinct from LGRBs (e.g. Berger 2010 and references therein), and should be considered separately in such studies.

Most recently, Han et al.\ (2010) compared the {\it M-Z} relation for SDSS galaxies from Liang et al.\ (2007) to a small sample of 5 LGRB host galaxies. While the sample size is small, and the comparison sample redshift is inhomogenous with the LGRB host redshifts, the LGRB host galaxies are found to consistently lie below the {\it M-Z} relation for SDSS galaxies.

In Figure 1 we plot the {\it M-Z} relation for our sample of $z < 1$ LGRB host galaxies. We find that these two parameters have a strong and statistically significant positive correlation (Pearson's $r$ = 0.80, $p$ = 0.001). This is a significant deviation from the results of Savaglio et al.\ (2009), who find no {\it M-Z} relation for their sample of GRB host galaxies. We postulate that this is primarily due to differences in metallicity determinations. While our stellar masses derived for these host galaxies are in agreement with the stellar masses of Savaglio et al.\ (2009), our metallicities are largely based on late-time spectroscopic observations from our ongoing host galaxy survey, and the LGRB host data plotted in Figure 1 are based on metallicities that were determined using the Kobulnicky \& Kewley (2004) R$_{23}$ calibration.

We have also compared our LGRB host galaxy data to two star-forming galaxy samples:

{\it SDSS galaxies}: For the nearby ($z < 0.3$) LGRB host galaxies, we adopt data from $\sim$53,000 star-forming SDSS galaxies as a comparison sample. The data plotted in Figure 1 are taken from Table 3 of Tremonti et al.\ (2004), and has been binned by mass in increments of $\sim$0.1 dex. The Tremonti et al.\ (2004) metallicities have been converted into the R$_{23}$ metallicity calibration of Kobulnicky \& Kewley (2004), using the conversion coefficients given in Table 3 of Kewley \& Ellison (2008). In addition, the Tremonti et al.\ (2004) stellar masses were derived using spectral indices, and Zahid et al.\ (2010) find that these masses differ from masses determined using the {\it Le Phare} code by a constant offset, attributable to the different IMFs and techniques (spectral vs. photometric) used in the two methods. As a result, we have decremented the Tremonti et al.\ (2004) stellar masses by the recommended offset of 0.17 dex to bring them into agreement with the stellar mass determinations of the {\it Le Phare} code; for more discussion see Zahid et al.\ (2010). The sample covers a redshift range of $0.005 < z < 0.25$, with a median redshift of $z \sim 0.1$.

{\it DEEP2 galaxies}: For the intermediate-redshift ($0.3 < z < 1$) LGRB host galaxies, we compare our results to stellar mass-binned data for 1,330 emission line galaxies from the Deep Extragalactic Evolutionary Probe 2 (DEEP2) survey. The stellar masses and metallicities for these galaxies were determined by Zahid et al.\ (2010), using the {\it Le Phare} stellar mass code and the Kobulnicky \& Kewley (2004) R$_{23}$ metallicity diagnostic (it is worth noting that the metallicities are based on equivalent width data rather than fluxes, which could introduce a systematic error of up to $\sim$0.5dex;  see Zahid et al.\ 2010). The data cover a redshift range from $0.75 < z < 0.82$. 

From this comparison, we find that most of the LGRB hosts in our sample fall below the standard {\it M-Z} relation for star-forming galaxies at similar redshifts, with differences ranging from $-0.05$ to $-0.75$ dex across a fixed stellar masses. We do note that, for the high-metallicity hosts of GRB 050826 and GRB 020819, the measured metallicities agree with the SDSS and DEEP2 {\it M-Z} relations to within the systematic errors. Across the whole sample we find an average offset from the general star-forming galaxy populations of $-0.42 \pm 0.18$ dex in metallicity ($-0.45 \pm$ 0.17 dex for the $z < 0.3$ sample, $-0.38 \pm$ 0.2 dex for the $0.3 < z < 1$ sample).

We must also consider the selection effects inherent in such a study. Host galaxy surveys are typically limited to LGRBs with well-detected optical afterglows that can be confidently associated with a host. As a result, these surveys are limited in their ability to sample ``dark" LGRBs; Fynbo et al.\ (2009) estimate an overall dark burst fraction of 25\% - 42\% based on their survey of 77 Swift LGRBs. The primary cause of the dark LGRB phenomenon remains unknown; however, recent evidence has supported the effects of dust extinction, in particular dust that is primarily present in the circumburst environment (Perley et al.\ 2009). Levesque et al.\ (2010b) suggest that this circumburst extinction could in turn be connected to high metallicity. A connection between dark LGRBs and higher-metallicity host environments is also discussed in Fynbo et al.\ (2009) and Graham et al.\ (2009). If the dark burst phenomenon is correlated with higher-metallicity host environments, and our sample is biased against the hosts of dark LGRBs, it is therefore possible that the apparent divergence of LGRBs from the general {\it M-Z} relation only holds true for the most nearby lower-mass sample. Future inclusion of additional dark burst host environments would help to further clarify the true nature of this observed LGRB host offset.

Finally, in Figure 1 we compare the {\it M-Z} relation determined for our intermediate-redshift LGRB hosts to binned data from the Erb et al.\ (2006) {\it M-Z} relation for a sample of 87 ultraviolet-selected star-forming galaxies at $z \gtrsim 2$. The metallicities for this sample, originally derived using the [NII]$\lambda$6584/H$\alpha$ diagnostic from Pettini \& Pagel (2004), have been converted to the Kobulnicky \& Kewley (2004) calibration according to  the coefficients in Table 3 of Kewley \& Ellison (2008); the masses are in agreement with the {\it Le Phare} code determination and do not require the offset decrement applied to the SDSS sample. Erb et al.\ (2006) find that their $z \sim 2$ sample is offset from the local {\it M-Z} relation by $\sim$0.3 dex; we find a smaller average offset of $\sim0.16$ dex between the Erb et al.\ (2006) sample and the Zahid et al.\ (2010) DEEP2 sample. This is $\sim$0.2 dex less than the average offset that we measure for our LGRB host sample at $0.3 < z < 1$. From this we can conclude that the low-metallicity offset seen here for LGRB host galaxies is smaller, though still present, when compared to star-forming galaxies at $z \sim 2$. Future observations of LGRB host galaxies out to $z \gtrsim 2$ are necessary to draw further conclusions about whether LGRB host galaxies may be useful tracers of the general star-forming galaxy population at higher redshifts.

\section{Discussion and Conclusions}
\label{disc}
In this paper we have presented a sample of 8 LGRB host galaxies, including 2 from the literature and 6 that have been observed as part of our ongoing rest-frame optical spectroscopic survey of this host population at $z < 1$. Using emission-line diagnostics we have determined metallicities, ionization parameters, young stellar population ages, and SFRs for these galaxies. In addition, we have combined the LGRB host galaxy data in this paper with LGRB host galaxies published in Paper I, determining stellar masses for these host galaxies and constructing a {\it M-Z} relation for LGRB host galaxies. We find a strong positive correlation between stellar mass and metallicity for LGRB host galaxies (Pearson's $r = 0.80$, $p = 0.001$), at odds with the previous results of Savaglio et al.\ (2009). From this {\it M-Z} relation, we have also concluded that LGRBs tend to occur in host galaxies with lower metallicities than the general population, and this that trend extends out to $z \sim 1$. However, this trend may become less pronounced at higher redshifts, where star-forming galaxy metallicities are lower on average (e.g., Kobulnicky \& Kewley 2004, Shapley et al.\ 2004, Erb et al.\ 2006, Chary et al.\ 2007, Dav\`{e} \& Oppenheimer 2007, Liu et al.\ 2008).

Levesque et al.\ (2010b) studied the host environment of GRB 020819, which was found to have an unusually high metallicity (log(O/H) + 12 = 9.0 from the [NII]/[OII] diagnostic of Kewley \& Dopita 2002). In this work we have now uncovered a second host galaxy, the $z = 0.296$ host of GRB 050826, with a high metallicity of log(O/H) + 12 = 8.83. GRB 050826 was a subluminous GRB, with a detected X-ray afterglow and optical transient (Mirabal et al.\ 2007). This host galaxy contradicts speculation in Levesque et al.\ (2010b) that high-metallicity host environments may be restricted to ``dark" LGRBs, with no detected optical afterglows (see also Graham et al.\ 2009). GRB 050826 is the first example of a ``classical" LGRB occurring in a host galaxy with such a high metallicity.  While it is true that several studies have measured high metallicities in other LGRB host galaxies based on afterglow spectra (e.g. Watson et al.\ 2006, El\'{i}asd\'{o}ttir et al.\ 2009, Prochaska et al.\ 2009), the relationship between afterglow absorption metallicities and emission-line metallicities has not yet been examined, and these values may not be directly comparable.

It is possible that the observed offset of our LGRB host galaxies from the general {\it M-Z} relation for star-forming galaxies could be attributable to the proposed relation between stellar mass, metallicity, and SFR described in Mannucci et al.\ (2010). An in-depth analysis of the Mannucci et al.\ (2010) relation and its potential application to the observed offset of our LGRB host sample is currently underway (Kewley et al.\ in prep).

The explanation behind the observed metallicity offset in the mass-metallicity relation impacts several intriguing questions regarding the role of metallicity in LGRB progenitors and host galaxies. While we have demonstrated that {\it most} LGRB host galaxies fall below the {\it M-Z} relation for the general galaxy population, we have also noted that the high-metallicity hosts of GRB 020819 and GRB 050826 do show agreement with the standard {\it M-Z} relations when considering the systematic errors. The metallicities of these two hosts also challenge the proposed belief that LGRBs follow a strict ``cut-off" metallicity for their host galaxies (Wolf \& Podsiadlowski 2007, Modjaz et al.\ 2008, Kocevski et al.\ 2009), and that a low metallicity is critical to generating a rapidly-rotating progenitor (e.g. Meynet \& Maeder 2005, Woosley \& Heger 2006, Yoon et al.\ 2006). In the case of GRB 020819, the specific explosion site was also found to have a high metallicity, demonstrating that the young progenitor itself evolved in a relatively metal-rich environment rather than a metal-poor pocket of the high-metallicity host and highlighting the importance of understanding the role that higher metallicities may play in the formation of dark bursts. Finally, the relativistic SN 2009bb (Soderberg et al.\ 2010) was also found to have a high-metallicity explosion site (Levesque et al.\ 2010c), further challenging the assumption that central-engine-driven relativistic explosions can only be produced by low-metallicity progenitors.

Combined, these results present a paradoxical conclusion. From this work and a number of recent host galaxy studies, it is clear that LGRBs occur preferentially in host galaxies with lower metallicities than the general star-forming galaxy population. However, we have also recently found that LGRBs do not require a low-metallicity progenitor environment, and that their host galaxies do not necessarily adhere to a strict low ``cut-off" metallicity. As a result, the physical mechanism that is driving this low-metallicity trend remains unclear. We must carefully examine how metallicity contributes to the progenitor production and explosive properties of LGRBs, and how this in turn might produce the observed trend towards host galaxies that fall below the general {\it M-Z} relation.

We thank the anonymous referee for valuable feedback and suggestions regarding this manuscript. We gratefully acknowledge useful correspondence with Andy Fruchter, John Graham, Olivier Ilbert, Sandra Savaglio, and Alicia Soderberg regarding this work. We are grateful for the hospitality and assistance of the W. M. Keck Observatories in Hawaii, in particular the guidance and assistance of Greg Wirth, Scott Dahm, and Jim Lyke, as well as the Las Campanas Observatory in Chile. This paper made use of data from the Gamma-Ray Burst Coordinates Network (GCN) circulars. The authors wish to recognize and acknowledge the very significant cultural role and reverence that the summit of Mauna Kea has always had within the indigenous Hawaiian community.  We are most fortunate to have the opportunity to conduct observations from this sacred mountain. E. Levesque's participation was made possible in part by a Ford Foundation Predoctoral Fellowship and an Einstein Fellowship. L. Kewley and E. Levesque gratefully acknowledge support by NSF EARLY CAREER AWARD AST07-48559. E. Berger acknowledges support by NASA/Swift AOJ grant 5080010.

\begin{deluxetable}{l c c c c c c c c}
\tabletypesize{\scriptsize}
\tablewidth{0pc}
\tablenum{1}
\label{tab:params} 
\tablecolumns{9}
\tablecaption{Keck LRIS Observing Set-ups}
\tablehead{
\colhead{Host Galaxy}
&\colhead{$\alpha_{2000}$}
&\colhead{$\delta_{2000}$}
&\colhead{Date (UT)}
&\colhead{Dichroic}
&\colhead{Grism}
&\colhead{Grating}
&\colhead{$\lambda_c$ (\AA)}
&\colhead{Exposure Time}
}
\startdata
GRB 980703   &23 59 06.72 &+08 35 07.08 &18 Nov 2009 &560 &\nodata &400/8500 &8500 &1800 $\times$ 6 \\
GRB 010921   &22 56 00.00 &+40 55 52.31 &19 Nov 2009 &680 &\nodata &400/8500 &8100 &1800 $\times$ 4 \\
GRB 020819\tablenotemark{a}   &23 27 19.44 &+06 15 55.80 &2 Nov 2008 &680 &300/5000 &400/8500 &8100 &1800 $\times$ 4 \\
GRB 070612A &08 05 29.61 &+37 16 15.20 &18 Nov 2009 &560 &\nodata &400/8500 &7000 &1800 $\times$ 6 \\
\enddata	  
\tablenotetext{a}{Observing set-up for the host nucleus, published in Levesque et al.\ (2010b).}    	      	
\end{deluxetable}

\begin{deluxetable}{l c c c c c c c c c}
\tabletypesize{\scriptsize}
\tablewidth{0pc}
\tablenum{2}
\label{tab:fluxes}
\tablecolumns{10}
\tablecaption{Diagnostic Emission-Line Fluxes\tablenotemark{a}}
\tablehead{
\colhead{Host Galaxy}
&\colhead{[O II] 3727}
&\colhead{H$\gamma$ 4340}
&\colhead{H$\beta$ 4861}
&\colhead{[O III] 4959}
&\colhead{[OIII] 5007}
&\colhead{H$\alpha$ 6563}
&\colhead{[N II] 6584}
&\colhead{[S II] 6717}
&\colhead{[S II] 6730}
}
\startdata
GRB 980703   &5.72   &1.09 &1.73 &1.60 &4.82 & \nodata & \nodata & \nodata & \nodata \\ 
GRB 991208 &0.44\tablenotemark{b} &0.12\tablenotemark{b} &0.34\tablenotemark{b} &0.24\tablenotemark{b} &0.57\tablenotemark{b} &1\tablenotemark{c} &0.05\tablenotemark{c} &\nodata &\nodata \\
GRB 010921   &2.74\tablenotemark{b} &\nodata &0.88\tablenotemark{b} &0.60\tablenotemark{b} &2.06\tablenotemark{b} &1\tablenotemark{c} &$<$0.04\tablenotemark{c} &\nodata &\nodata \\ 
GRB 020819   &2.41\tablenotemark{d} &\nodata &1.66\tablenotemark{d} &\nodata &0.86\tablenotemark{d} &9.62\tablenotemark{d} &4.10\tablenotemark{d} &\nodata &\nodata \\ 
GRB 050826\tablenotemark{e}   &1.13 &\nodata &0.64 &0.29 &0.86 &3.30 &\nodata &\nodata &\nodata \\ 
GRB 050826\tablenotemark{f} &\nodata &\nodata &\nodata &\nodata &\nodata &1\tablenotemark{c} &0.17\tablenotemark{c} &0.16\tablenotemark{c} &0.10\tablenotemark{c} \\
GRB 070612A &7.31 &1.12 &3.36 &\nodata &3.78 &1\tablenotemark{c} &$<$0.01\tablenotemark{c} &\nodata &\nodata 
\enddata	      
\tablenotetext{a} {Raw measured fluxes in units of $10^{-16}$ ergs cm$^2$ s$^{-1}$ \AA$^{-1}$}   
\tablenotetext{b} {From Paper I.} 
\tablenotetext{c} {Relative flux, normalized to H$\alpha$.} 
\tablenotetext{d} {From Levesque et al.\ (2010b) observations of the host galaxy nucleus.}
\tablenotetext{e} {Observations from 14 Jan 2008.}
\tablenotetext{f} {Observations from 6 Jan 2006.}
	
\end{deluxetable}

\clearpage
\LongTables
\begin{landscape}
\begin{deluxetable}{l c c c c c c c c c c}
\tabletypesize{\scriptsize}
\tablewidth{0pc}
\tablenum{3}
\label{tab:gals} 
\tablecolumns{11}
\tablecaption{ISM Properties of LGRB Host Galaxies}
\tablehead{
\colhead{Galaxy}
&\colhead{$z$}
&\multicolumn{2}{c}{log(O/H) + 12\tablenotemark{a}}
&\colhead{log($q$)}
&\colhead{E($B-V$)\tablenotemark{b}}
&\colhead{W$_{H\beta}$\tablenotemark{c}}
&\colhead{Age (Myr)\tablenotemark{d}}
&\colhead{$M_B$ (mag)\tablenotemark{e}}
&\colhead{SFR (M$_{\odot}$/yr)}
&\colhead{log($M_{\star}/M_{\odot}$)}\\ \cline{3-4}
\multicolumn{2}{c}{}
&\colhead{R$_{23}$}
&\colhead{PP04}
&\multicolumn{6}{c}{}
&\colhead{(M$_{\odot}$)}
}
\startdata
{\bf This Paper} & & & & & & & & & & \\
\hline
GRB 980703 &0.966  &8.31/8.65 &\nodata &7.51/7.66 &0.00 &90.5 &4.7 $\pm$ 0.1/4.4 $\pm$ 0.2 &-21.4 &9.9/13.6\tablenotemark{f} &9.83 $\pm$ 0.13 \\ 
GRB 991208 &0.706  &8.02 &\nodata  &7.38 &0.58                                 &99.8    &4.2 $\pm$ 0.2 &-18.5 &3.47\tablenotemark{f} &8.85 $\pm$ 0.17 \\ 
GRB 010921\tablenotemark{g} &0.451  &8.24 &\nodata  &7.44 &0.00                                 &11.7    &8.0 $\pm$ 0.2 &-19.4 &0.70\tablenotemark{f} &9.56$^{+0.09}_{-0.11}$ \\ 
GRB 020819\tablenotemark{h} &0.410  &9.0 &8.8 &\nodata &0.71 &5.08 &7.8 $\pm$ 0.9 &\nodata &23.6 &10.65 $\pm$ 0.19 \\ 
GRB 030528 &0.782  &$\sim$8.40 &\nodata &\nodata &$>$0.46\tablenotemark{i} &\nodata &\nodata &-20.53 &$>$12.1\tablenotemark{f} &9.11$^{+0.23}_{-0.26}$ \\ 
GRB 050824 &0.828  &$\sim$8.40 &\nodata &\nodata &$<$0.16\tablenotemark{j} &\nodata &\nodata &\nodata &$<$0.941\tablenotemark{f} &\nodata \\  
GRB 050826 &0.296  &8.83 &\nodata &7.51 &0.60 &20.1 &5.9 $\pm$ 0.7 &-19.7 &2.94 &10.10$^{+0.22}_{-0.26}$ \\ 
GRB 070612A &0.671  &8.29 &\nodata &7.28 &0.64 &30.53 &5.8 $\pm$ 0.2 &\nodata &81\tablenotemark{f} &\nodata \\ 
\hline 
{\bf Paper I} & & & & & & & & & & \\
\hline
GRB 980425 &0.009  &$\sim$8.40    &8.28        &\nodata    &0.34 &\nodata         &$\sim$5.0\tablenotemark{k} &-17.6 &0.57 &9.22 $\pm$ 0.52 \\ 
GRB 990712 &0.434  &$\sim$8.40 &\nodata  &\nodata &0.57                                 &\nodata &\nodata &-18.6            &10.7\tablenotemark{f} &9.15 $\pm$ 0.04 \\ 
GRB 020405 &0.691  &8.33/8.59 &\nodata  &7.65/7.78 &0.00                                 &25.6   &6.2 $\pm$ 0.2/5.4 $\pm$ 0.3 &\nodata  &1.61/2.05\tablenotemark{f} &\nodata \\ 
GRB 020903 &0.251  &8.07          &7.98        &8.15          &0.00                                &31.3   &5.8 $\pm$ 0.2 &-18.8  &1.7 &8.79$^{+0.19}_{-0.24}$ \\ 
GRB 031203\tablenotemark{l} &0.105 &8.27          &8.10        &8.37         &1.17                                 &103.9    &4.7 $\pm$ 0.1 &-21.0 &4.8 &8.26 $\pm$ 0.45 \\ 
GRB 030329 &0.168 &8.13          &8.00        &7.80         &0.13                                 &59.6    &4.9 $\pm$ 0.1 &-16.5 &1.2 &7.91$^{+0.12}_{-0.44}$ \\
GRB 051022 &0.807  &8.62          &8.37        &7.55         &0.50                                 &29.0    &5.2 $\pm$ 0.3 &-21.8  &271\tablenotemark{f} &10.42 $\pm$ 0.05 \\ 
GRB 060218 &0.034 &8.21          &8.07        &7.71         &0.01                                 &33.2   &5.7 $\pm$ 0.2 &-15.9 &0.03 &8.37 $\pm$ 0.14 \\
\enddata	     
\tablenotetext{a}{Metallicities have a systematic error of $\pm$0.1 dex due to uncertainties in the strong line diagnostics (Kewley \& Dopita 2002).} 	
\tablenotetext{b}{Total color excess in the direction of the galaxy, used to correct for the effects of both Galactic and intrinsic extinction.}
\tablenotetext{c}{Rest-frame equivalent widths.}
\tablenotetext{d}{Ages come from the equations derived for the  Schaerer \& Vacca (1998) models relating H$\beta$ equivalent widths and galaxy ages, adopting the R$_{23}$ metallicities (Paper I).}
\tablenotetext{e} {$M_B$ values come from the literature as follows: Hammer et al.\ 2006 (GRB 980425), Christensen et al.\ 2004 (GRB 980703, GRB 990712, GRB 991208, GRB 010921), Soderberg et al.\ 2004 (GRB 020903), Rau et al.\ 2005 (GRB 030528), Margutti et al.\ 2007 (GRB 031203), Gorosabel et al.\ 2005 (GRB 030329), Castro-Tirado et al.\ 2007 (GRB 051022), Mirabal et al.\ 2007 (GRB 050826), and Wiersema et al.\ 2007 (GRB 060218).}
\tablenotetext{f} {SFR determined from the [OII] line flux and the metallicity-dependent relation from Kewley et al.\ (2004).}
\tablenotetext{g} {GRB 010921 was originally published in Paper I; however, its physical properties have been updated in this paper.}
\tablenotetext{h} {GRB 020819 values are from observations of the host nucleus in Levesque et al.\ (2010b).}
\tablenotetext{i} {Lower limit from Rau et al.\ (2005).}
\tablenotetext{j} {Upper limit from Sollerman et al.\ (2007).}
\tablenotetext{k} {Value from Christensen et al.\ (2008).}
\tablenotetext{l} {Since the host of GRB 031203 is not classified as a purely star-forming galaxy, all ISM properties should be taken as approximate, given the potential unknown contribution of AGN activity.}
\end{deluxetable}
\clearpage
\end{landscape}

\begin{figure}
\epsscale{1}
\plotone{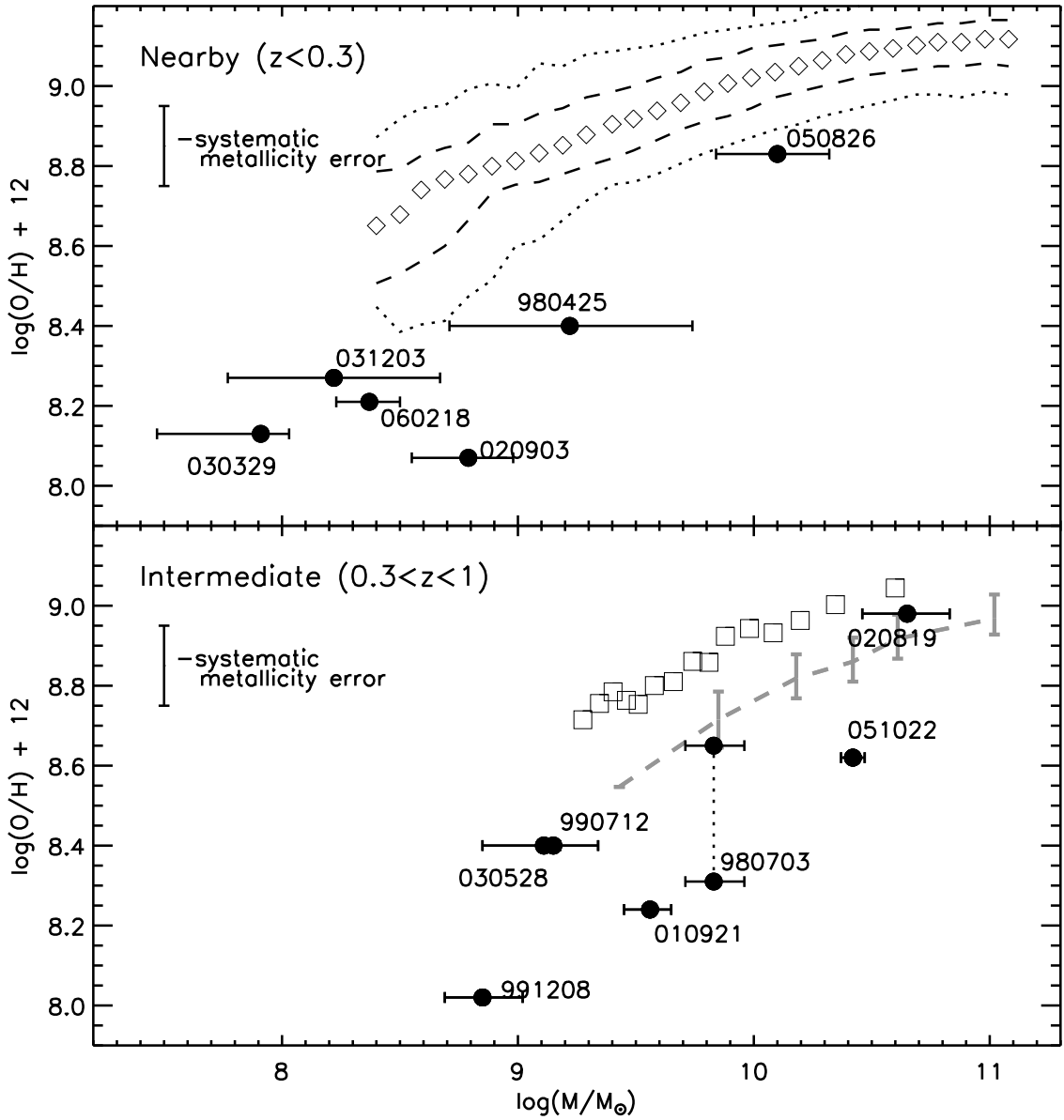}
\caption{The mass-metallicity relation for both nearby ($z < 0.3$, top) and intermediate-redshift ($0.3 < z < 1$, bottom) LGRB host galaxies (filled circles). We compare the nearby LGRB hosts to the binned mass-metallicity data from Tremonti et al.\ (2004) for a sample of $\sim$53,000 star-forming SDSS galaxies, where the open diamonds represent the median in each bin, and the dashed and dotted lines show the contours which include 68\% and 95\% of the data, respectively. For the intermediate-redshift LGRB hosts, we plot binned mass-metallicity data for a sample of 940 emission line galaxies from the DEEP2 survey (Zahid et al.\ 2010; open squares). All metallicities correspond to the Kobulnicky \& Kewley (2004) R$_{23}$ diagnostic. For the $z = 0.966$ host galaxy of GRB 980703, where we cannot distinguish between the lower and upper branches of the R$_{23}$ diagnostic, we plot both metallicities and connect the resulting data points with a dotted line to indicate their origin from a single host spectrum. The Erb et al.\ (2006) {\it M-Z} relation at $z \sim 2$ is plotted against our intermediate-redshift data as a gray dashed line.}
\end{figure}

\clearpage
\appendix
\section{LGRB Host Galaxy Properties}
\subsection{GRB 980703} 
The host galaxy of GRB 980703 is at an intermediate redshift of $z = 0.966$ (Figure A1). The H$\gamma$/H$\beta$ ratio in this galaxy gives us an E($B-V$) = 0, and thus no correction for extinction is applied. We can apply the Kobulnicky \& Kewley (2004) R$_{23}$ metallicity diagnostic to this galaxy, but without a detection of the H$\alpha$ and [NII]$\lambda$6584 features we cannot determine whether it lies on the lower or upper branches of the diagnostic. Calculating metallicities for both branches, we find log(O/H) + 12 = 8.31 $\pm$ 0.1 (lower; log $q$ = 7.51) and log(O/H) + 12 = 8.65 $\pm$ 0.1 (upper; log $q$ = 7.66). We also determine a young stellar population age for this host galaxy of 4.7 $\pm$ 0.1 Myr for the lower-branch metallicity and 4.4 $\pm$ 0.2 Myr for the upper-branch metallicity. Using the flux of the [OII]$\lambda$3727 line and the Kewley et al.\ (2004) metallicity-dependent relation, we determine SFRs of 9.9 $M_{\odot}$ yr$^{-1}$ for the lower-branch metallicity and 13.6$M_{\odot}$ yr$^{-1}$ for the upper-branch metallicity. These SFRs agree with the lower limit of $>7 M_{\odot}$ yr$^{-1}$ determined by Djorgovski et al.\ (1998) and the 8-13 $M_{\odot}$ yr$^{-1}$ range found by Holland et al.\ (2001). Finally, with photometry from Savaglio et al.\ (2009) and the {\it Le Phare} code we find a stellar mass for the host galaxy of log($M_{\star}/M_{\odot}$) = 9.83 $\pm$ 0.13.

\subsection{GRB 991208}
The host galax of GRB 991208 is an intermediate-redshift ($z = 0.706$) host galaxy that was originally published in Paper I. We previously applied the Kobulnicky \& Kewley (2004) R$_{23}$ metallicity diagnostic to our LRIS observations of this host, but were unable to established whether the host metallicity was on the lower or upper branch of the double-valued diagnostic. Here we present our NIRSPEC data for this host, which show a detection of the H$\alpha$ emission feature and an upper limit on the [NII]$\lambda$6584 feature (Figure A2). Based on the [NII]/H$\alpha$ ratio determined from this data, and following the criteria of Kewley \& Ellison (2008), we can now conclude that the host galaxy of GRB 991208 falls on the lower branch of the R$_{23}$ diagnostic, yielding a host metallicity of log(O/H) + 12 = 8.02 and an ionization parameter of log $q$ = 7.38 according to Kobulnicky \& Kewley (2004). We also derive a young stellar population age of 4.2 $\pm$ 0.2 Myr, and a metallicity-dependent SFR = 3.47 $M_{\odot}$ yr$^{-1}$ based on the Kewley et al.\ (2004) [OII] diagnostic. Using photometry from Savaglio et al.\ (2009) and the {\it Le Phare} code, we determine a stellar mass for this host galaxy of log($M_{\star}/M_{\odot}$) = 8.85 $\pm$ 0.17. 

\subsection{GRB 010921} 
The host galaxy of GRB 010921 is an intermediate-redshift ($z = 0.451$) host that was previously examined in Paper I. We originally applied the Kobulnicky \& Kewley (2004) R$_{23}$ metallicity diagnostic to these host observations, but were unable to determine whether this host was on the lower or upper branch of the diagnostic. Here we present our observation of the H$\alpha$ emission feature and an upper limit on the [NII]$\lambda$6584 feature (Figure A3). Based on the emission-line ratio determined from this observation and the criteria of Kewley \& Ellison (2008), we can now conclude that the host of GRB 010921 lies on the lower branch of the R$_{23}$ diagnostic, with a metallicity of log(O/H) + 12 = 8.24 $\pm$ 0.1, an ionization parameter of log $q$ = 7.44, a young stellar population age of 8.0 $\pm$ 0.2 Myr, and a SFR = 0.70 $M_{\odot}$ yr$^{-1}$ based on the metallicity-dependent [OII] diagnostic of Kewley et al.\ (2004). Adopting photometry from Savaglio et al.\ (2009) and using the {\it Le Phare} code, we also determine a stellar mass for the host galaxy of log($M_{\star}/M_{\odot}$) = 9.56$^{+0.09}_{-0.11}$.

\subsection{GRB 020819} 
For a detailed discussion of the unusual host galaxy of GRB 020819, see Levesque et al.\ (2010b); for this work we adopt the ISM properties derived for the nucleus of the host galaxy. We adopt photometry from Savaglio et al.\ (2009) and use the {\it Le Phare} code to determine a stellar mass for the host galaxy of log($M_{\star}/M_{\odot}$) = 10.65 $\pm$ 0.19.

\subsection{GRB 030528} 
Rau et al.\ (2005) publish emission-line fluxes, uncorrected for extinction, for the [OII]$\lambda$3727, H$\beta$, [OIII]$\lambda$4959, and [OIII]$\lambda$5007 features in the $z = 0.782$ host galaxy of GRB 030528. They also include upper limits on the [NeIII]$\lambda$3869, H$\delta$, and H$\gamma$ emission features. Rau et al.\ (2005) propose a total line-of-sight $A_V < 2.5$ for this host, corresponding to a Galactic E($B-V$) $<$ 0.62 from Schlegel et al.\ (1998) and an additional host extinction of E($B-V$) $<$ 0.19. However, Dutra et al.\ (2003) suggest a lower line-of-sight E($B-V$) = 0.46, following a rescaling of the Schlegel et al.\ (1998) extinction. We consider both of these proposed E($B-V$) values in our analysis, and find that in both cases the R$_{23}$ value places the host metallicity on the log(O/H) + 12 $\sim$ 8.4 $\pm$ 0.1 turnover of the Kobulnicky \& Kewley (2004) diagnostic. We also find a lower limit of SFR $> 12.1 M_{\odot}$ yr$^{-1}$ based on the metallicity-dependent [OII] relation of Kewley et al.\ (2004). Using photometry from Savaglio et al.\ (2009) and the {\it Le Phare} code, we find a stellar mass for the host galaxy of log($M_{\star}/M_{\odot}$) = $9.11^{+0.23}_{-0.26}$.

\subsection{GRB 050824}
Sollerman et al.\ (2007) publish fluxes for the [OII]$\lambda$3727, [NeIII]$\lambda$3869, H$\beta$, [OIII]$\lambda$4959, and [OIII]$\lambda$5007 emission features in the $z = 0.828$ host galaxy of GRB 050824. These fluxes are uncorrected for Galactic extinction (E($B-V$) = 0.035 from Schlegel et al.\ 1998) or host extinction; however, they estimate a host extinction of E($B-V$) $<$ 0.16, which we adopt here. Sollerman et al.\ (2007) find log(R$_{23}$) $\sim$ 1, which corresponds to the turnover of the Kobulnicky \& Kewley (2004) R$_{23}$ metallicity diagnostic; we find the same result, and determine a metallicity for the host galaxy of log(O/H) + 12 $\sim$ 8.4 $\pm$ 0.1, which remains unchanged across the full range of $E(B-V)$. We also measure SFR $< 0.941 M_{\odot}$ yr$^{-1}$ for the host using the metallicity-dependent relation for [OII] from Kewley et al.\ (2004), slightly lower than the Sollerman et al.\ (2007) value of $1.8 M_{\odot}$ yr$^{-1}$ using the Kennicutt (1998) [OII] and H$\alpha$ relations.

\subsection{GRB 050826}
The host of GRB 050826 is a low-redshift galaxy in our sample at $z = 0.296$ (Figure A4). Based on the H$\alpha$ and H$\beta$ line fluxes we observe for this host, we determine a total line-of-sight E($B-V$) = 0.60, which we adopt when correcting for extinction. We use the [NII]/H$\alpha$ ratio to place this galaxy on the upper branch of the Kobulnicky \& Kewley (2004) R$_{23}$ metallicity diagnostic; this gives us a surprisingly high log(O/H) + 12 = 8.83 $\pm$ 0.1 (log $q$ = 7.51). We determine a young stellar population age for this host of 5.9 $\pm$ 0.7 Myr. We use the H$\alpha$ flux and the Kennicutt (1998) relation to determine SFR = 2.94 $M_{\odot}$ yr$^{-1}$. Using the photometry of Savaglio et al.\ (2009) and the {\it Le Phare} code, we determine a stellar mass for the host galaxy of log($M_{\star}/M_{\odot}$) = 10.10$^{+0.22}_{-0.26}$.

\subsection{GRB 070612A}
The host galaxy of GRB 070612A is at an intermediate redshift of $z = 0.671$. Based on the H$\beta$ and H$\gamma$ line fluxes observed in this host (Figure A5, top), we determine a total line-of-sight E($B-V$) = 0.64, which we adopt when correcting our observed line fluxes for extinction. Using the H$\alpha$ detection and [NII]$\lambda$6584 upper limit determined from our NIRSPEC observations of the host (Figure A5, bottom), along with the criteria of Kewley \& Ellision (2008), we place this galaxy on the lower branch of the Kobulnicky \& Kewley (2004) R$_{23}$ metallicity diagnostic, with log(O/H) + 12 = 8.29 $\pm$ 0.1 (log $q$ = 7.28). We also determine a young stellar population age of 5.8 $\pm$ 0.2 Myr for the host, and a SFR = 81 $M_{\odot}$ yr$^{-1}$ based on the [OII] flux and the metallicity-dependent relation of Kewley et al.\ (2004). 

\clearpage
\begin{figure}
\epsscale{1}
\plotone{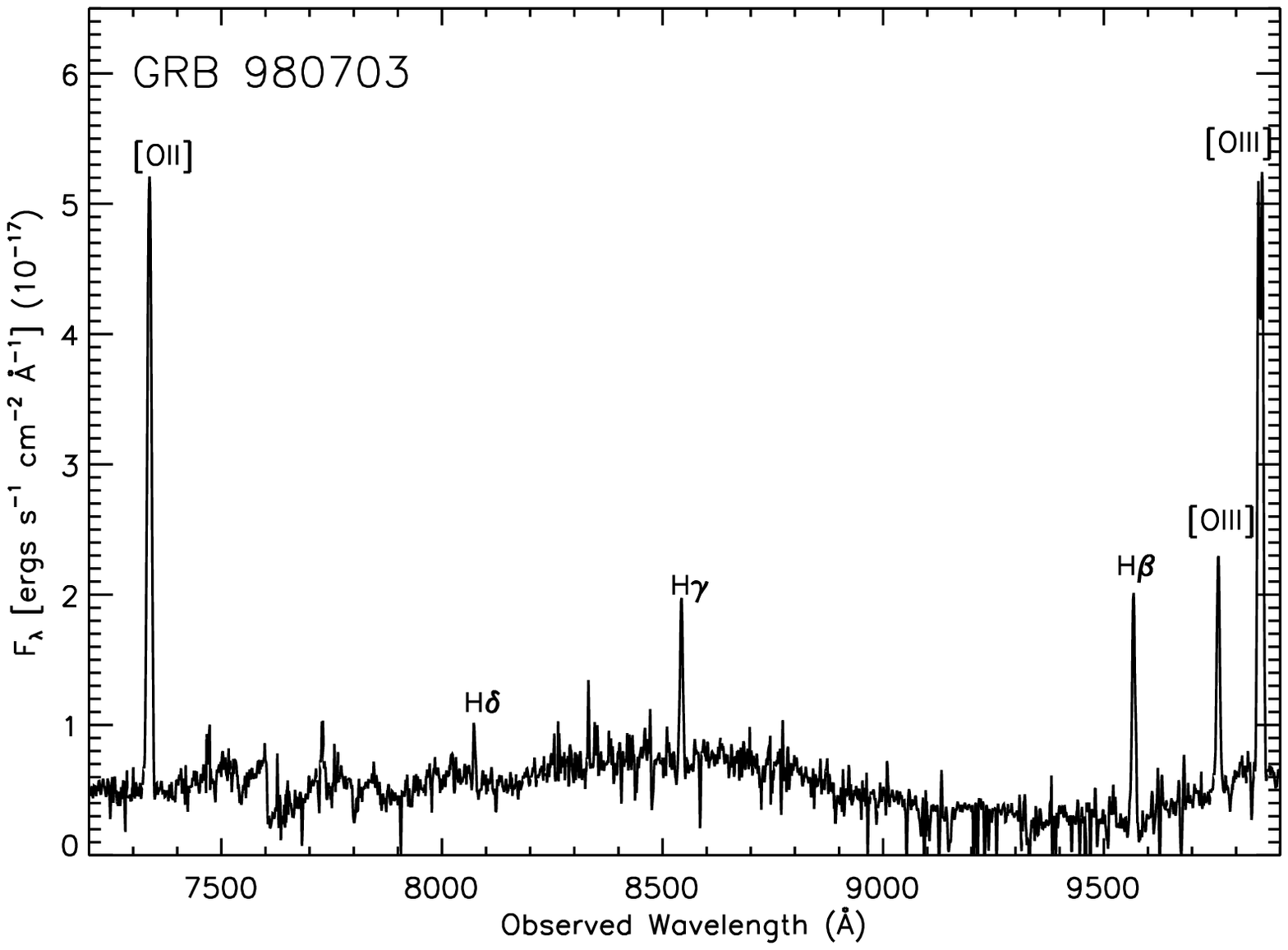}
\caption{Our spectrum of the $z = 0.966$ host galaxy of GRB 980703, observed with LRIS at Keck I on 18 November 2009.}
\end{figure}

\begin{figure}
\plotone{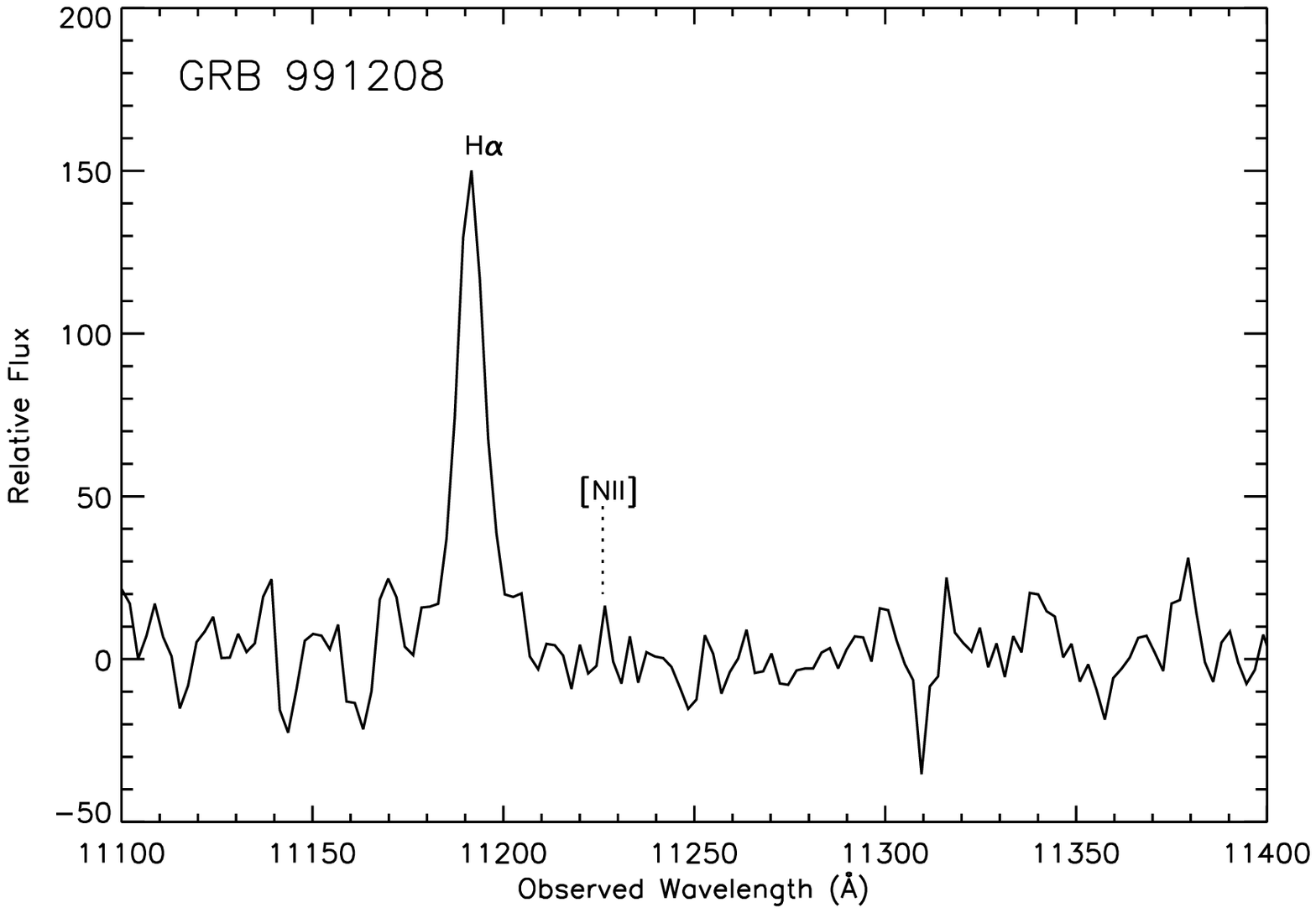}
\caption{Our spectrum of the H$\alpha$ feature and [NII]$\lambda$6584 upper limit in the $z = 0.706$ host galaxy of GRB 991208, observed with NIRSPEC at Keck II on 2 May 2010.}
\end{figure}

\begin{figure}
\plotone{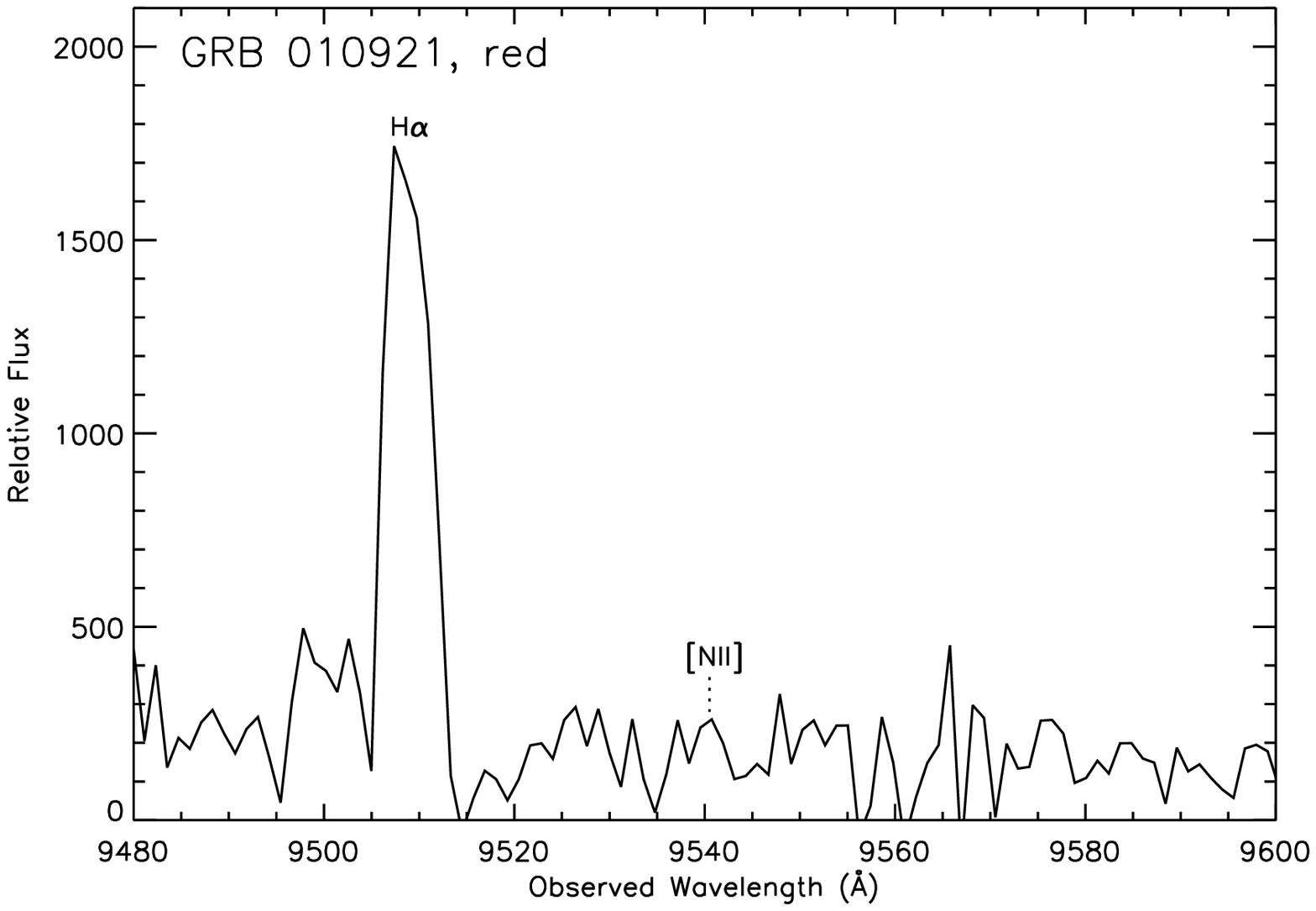}
\caption{Our spectrum of the H$\alpha$ feature and [NII]$\lambda$6584 upper limit in the $z = 0.451$ host galaxy of GRB 010921, observed with LRIS at Keck I on 19 November 2009.}
\end{figure}

\begin{figure}
\plotone{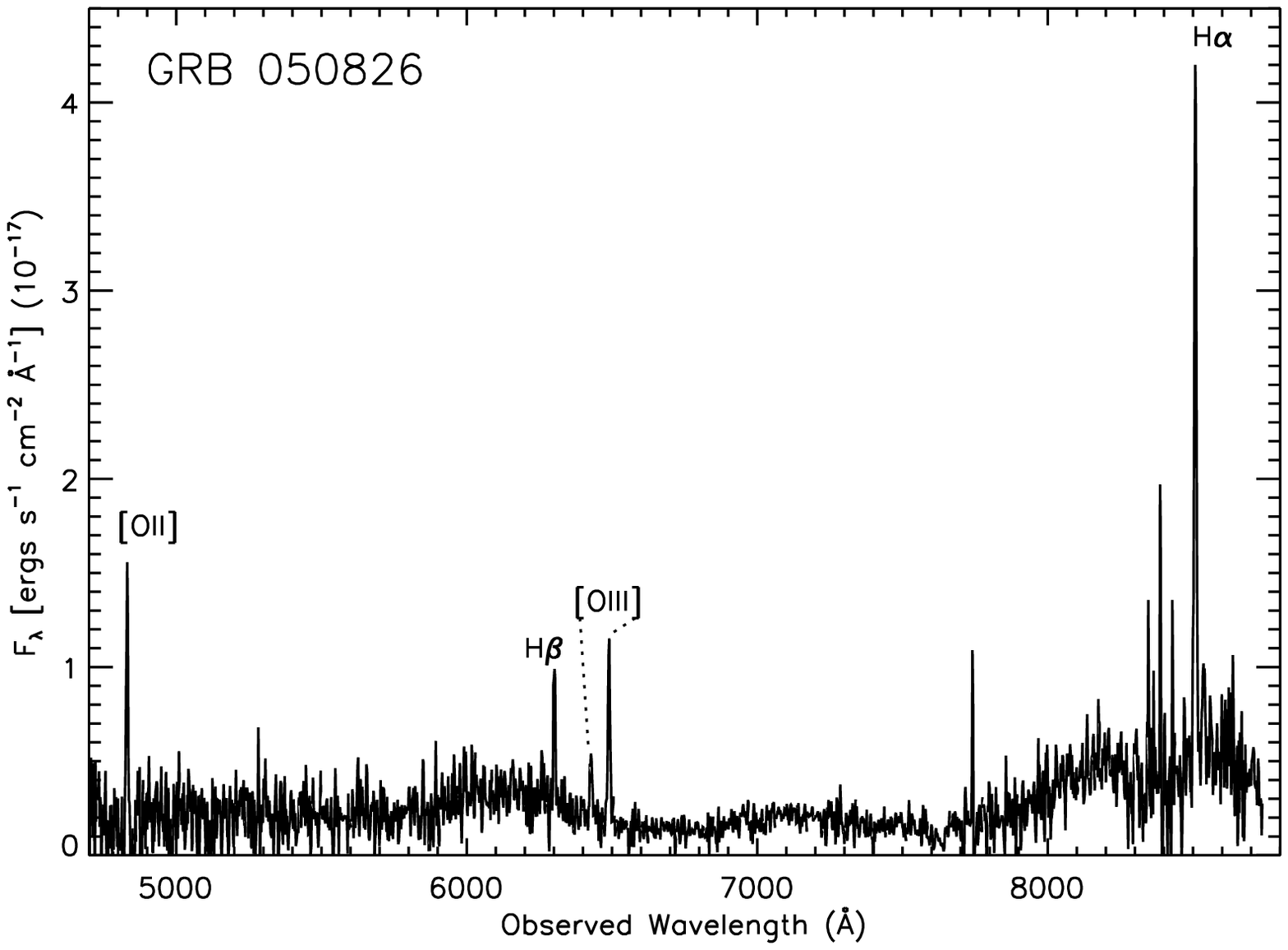}
\plotone{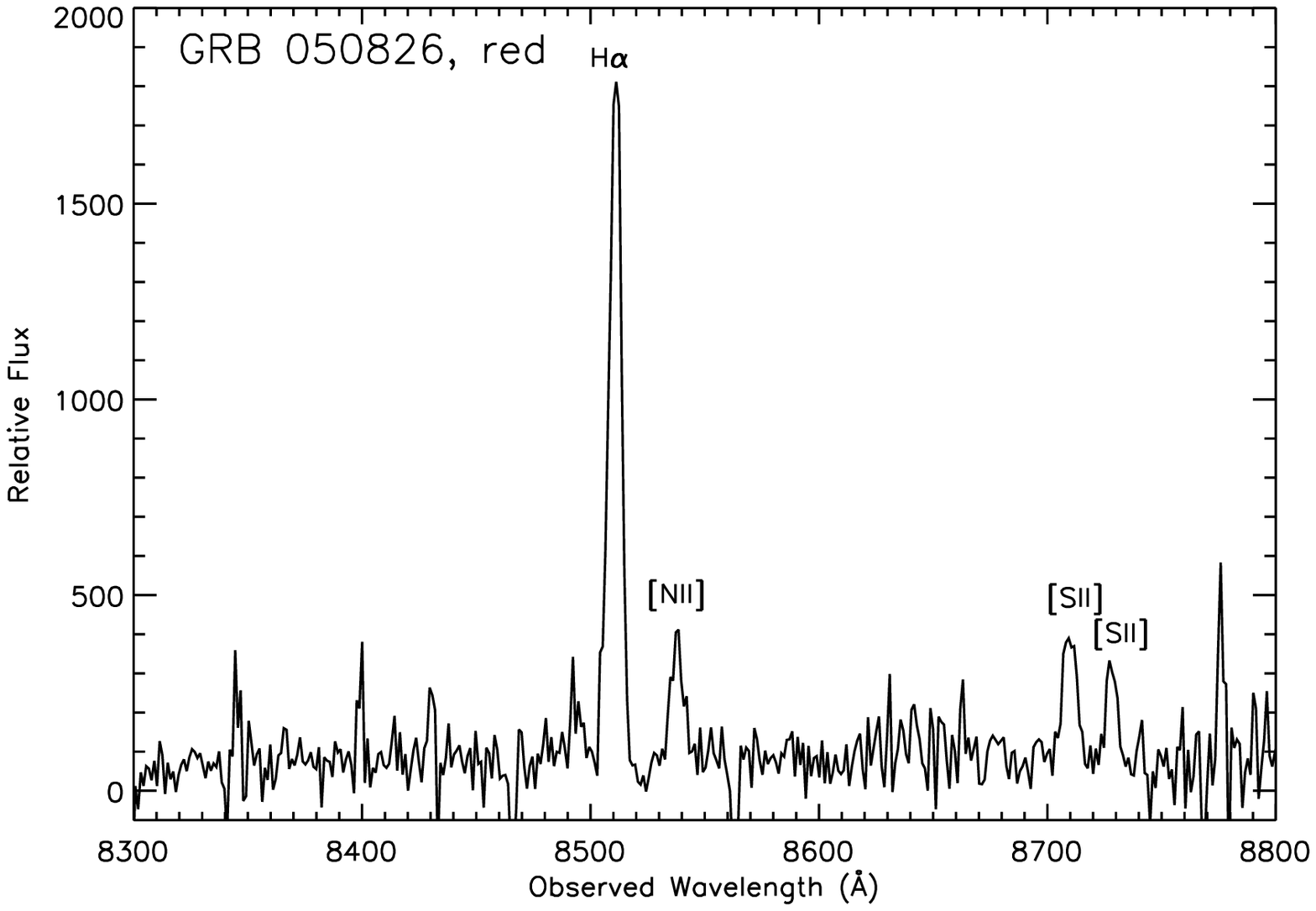}
\caption{Our spectra of the $z = 0.296$ host galaxy of GRB 050826, observed with LDSS3 at Magellan on 14 January 2008 (top) and 6 January 2006 (bottom). The 2008 spectrum covers the full range of spectral features; the 2006 spectrum shows detections of the H$\alpha$, [NII] $\lambda$6584, and [SII] $\lambda\lambda$6717,6731 emission features.}
\end{figure}

\begin{figure}
\plotone{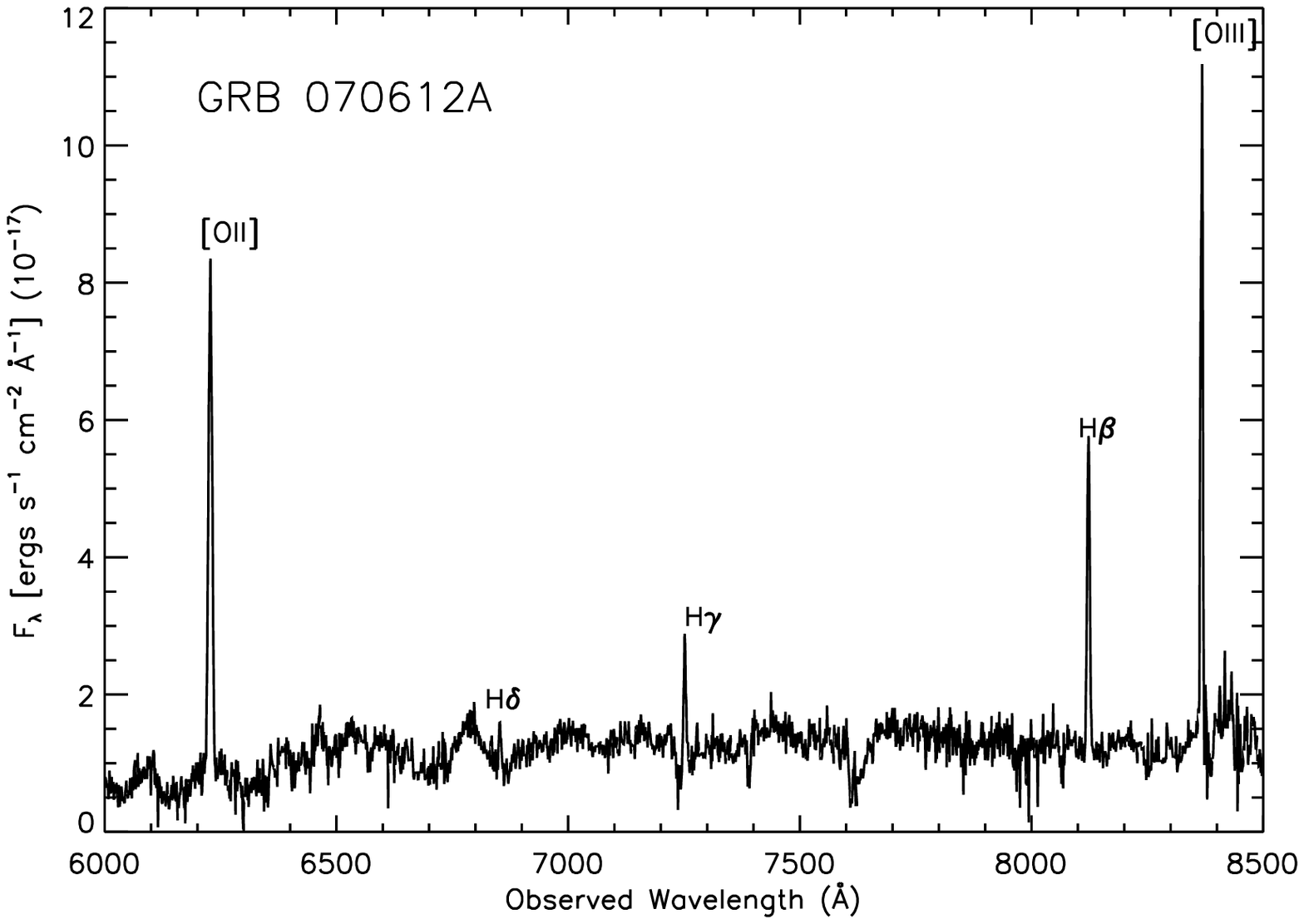}
\plotone{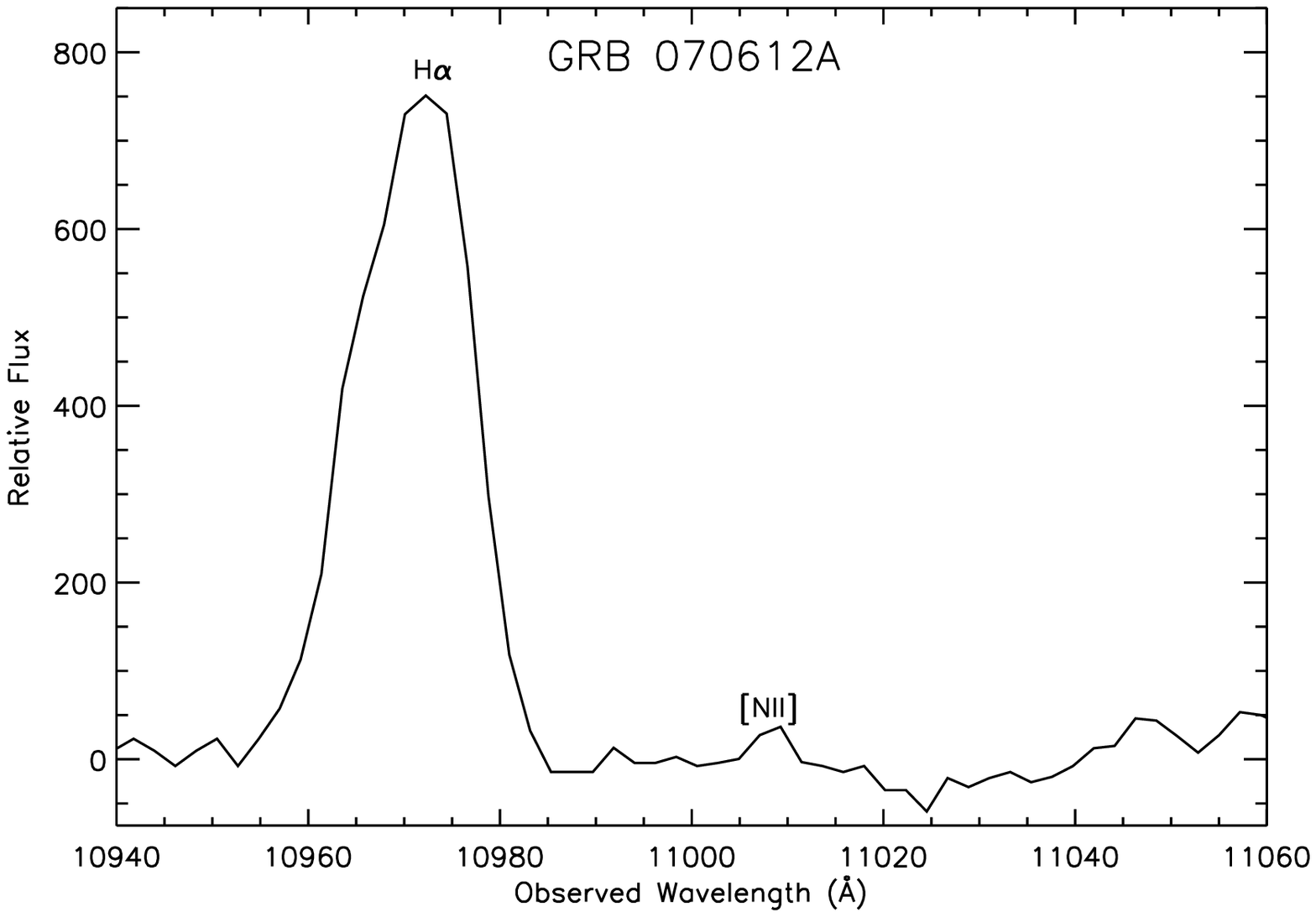}
\caption{Our spectra of the $z = 0.671$ host galaxy of GRB 070612A, obseved with LRIS at Keck I on 18 November 2009 (top; [OII] $\lambda$3727, H$\delta$, H$\gamma$, H$\beta$, and [OIII] $\lambda$5007) and with NIRSPEC at Keck II on 3 November 2009 (bottom; H$\alpha$ and our upper limit for [NII] $\lambda$6584).}
\end{figure}

\end{document}